# Energy backflow in unidirectional monochromatic and space-time waves

PEETER SAARI[1,2,*] AND IOANNIS BESIERIS[3]

[1]*Institute of Physics, University of Tartu, W. Ostwaldi 1, 50411, Tartu, Estonia*
[2]*Estonian Academy of Sciences, Kohtu 6, 10130 Tallinn, Estonia*
[3]*The Bradley Department of Electrical and Computer Engineering, Virginia Polytechnic Institute and State University, Blacksburg, Virginia 24060, USA*
[*]*peeter.saari@ut.ee*

**Abstract:** Backflow, or retropropagation, is a counterintuitive phenomenon whereby for a forward-propagating wave the energy locally propagates backward. In the context of backflow, physically most interesting are the so-called unidirectional waves, which contain only forward-propagating plane wave constituents. Yet, very few such waves possessing closed-form analytic expressions for evaluation of the Poynting vector are known. In this study, we examine energy backflow in a novel (2+time)-dimensional unidirectional monochromatic wave and in a (2+1)D spatiotemporal wavepacket, analytic expressions which we succeeded to find. We also present a detailed study of the backflow in the "needle" pulse. This is an interesting model object because well-known superluminal non-diffracting space–time wave packets can be derived from its simple factored wave function. Finally, we study the backflow in an unidirectional version of the so-called focus wave mode—a pulse propagating luminally and without spread, which is the first and most studied representative of the (3+1)D non-diffracting space–time wave packets (also referred to as spatiotemporally localized waves).

## Contents

## 1. Introduction

The phenomenon of backflow takes place when some quantity (energy density flow, local momentum, probability flow, etc.) in some spatio-temporal region of a wavefield is directed backward with respect to the directions of all plane-wave constituents of the wavefield. Backflow is a wave phenomenon that may occur in all kinds of wavefields describable by different types of wave equations [1]. Position probability backflow specific to quantum particles, such as electrons, has been termed 'quantum backflow,' and this subject is actively studied (see a review [2] and references therein). Recently, the similarities and differences between quantum backflow and backflow in classical wavefields have been discussed intensively [1, 3–5].

It has been known for quite a long time that in a superposition of four appropriately polarized and directed electromagnetic plane waves the Poynting vector direction can be reversed with respect to the direction of propagation of the resultant wave and its constituents [6]. A rather prominent energy backflow exhibited in such a quartet of electromagnetic waves was investigated recently in detail [7, 8]. A very recent study demonstrated that unbounded regions of reverse energy flow can be achieved with just two point light sources [9].

In the physical optics community, the energy backflow in sharply focused light has been known for more than half a century, and has been thoroughly studied theoretically recently [10–13]. Although our study deals solely with free-space fields, we note that the backflow has been studied also in dielectric microspheres [14], in evanescent fields and superoscillatory hotspots [15], and in the near field area of light scattered by nanoparticles and nanowires [16, 17]. In the context of quantum backflow, monochromatic optical fields have been used for recent experimental verification of the effect [18, 19]. Energy backflow in electromagnetic Bessel beams has been analytically demonstrated in [20, 21] and also in pulsed electromagnetic X waves [22]. Important is the circumstance that the backflow appears only in certain polarization geometries and occurs neither in scalar Bessel beams nor in a scalar X wave of zeroth order. The latter belongs to the class of the so-called nondiffracting (nonspreading, propagation-invariant) localized waves (LWs)—also known as space–time wave packets (STWP), especially their two-dimensional versions—which have been studied intensively during the past thirty years not only theoretically but also experimentally (see, e.g., [23–36]). They constitute spatiotemporally localized solutions to various hyperbolic equations governing acoustic, electromagnetic and quantum wave phenomena and can be classified according to their group velocity as luminal LWs, or focus wave modes (FWMs), superluminal LWs, or X waves, and subluminal LWs. Overviews of theoretical and experimental studies can be found in two edited monographs on the subject [37, 38] and in the recent thorough review article [39].

From the point of view of the backflow phenomenon, it is principally crucial whether the plane-wave constituents of the wave packet propagate only in the positive, say $z$-direction, or also backward. In other words, whether the support of the spectrum in the wave vector space is restricted to the positive half-space only of the $z$-component of the wave vector. In this case, the wave—we call it unidirectional—can be launched from an aperture as a freely propagating beam and the very question of energy backflow is meaningful. It must be pointed out that, in general, the group velocity in a wave packet typically differs from the energy velocity; see [40, 41].

In [42], some exact localized solutions to the wave equation which are totally free of backward components were found. In [43], a simple analytic expression for a unidirectional propagation-*variant* finite-energy pulse was determined. However, occurrence of possible backflow was not considered in these papers. Using as an example of a particular version of the expression of a hopfion-like unidirectional pulse, backflow was shown explicitly in [1]. In our work [44], we studied the backflow in scalar-valued and vector-valued generalizations of this propagation-*variant* (focusing) pulse. Additional generalizations of the unidirectional propagation-variant finite-energy pulses have been introduced by Lekner in [45]. Ref. [46] shows how analytical proof of existence of the backflow may be tricky. Therefore, in the present study, we focus on numerical–graphical methods.

The aim of this paper is to collect new data about conditions and locations of emergence of the backflow effect in a wavefield, and about dependence of characteristics of the effect on the type and parameters of wavefields. We hope that together with previous results on the subject, our study will deepen the understanding of physics of the energy backflow.

The paper is organized as follows. After a brief description in the next section of the methods used, in Section 3.1, we introduce a new solution to the Helmholtz equation in two dimensions—an exceptional successful result of our search for closed-form expressions of unidirectional monochromatic wavefields. The backflow in the corresponding harmonically time-dependent scalar and vector-valued wavefields are studied in Section 3.1.2. In Section 3.2, we introduce a novel wave packet formed from the monochromatic constituents considered in the previous subsection and study the backflow in the packet which constitutes two pulse fronts at positive times running away from the symmetry axis. Section 3.3 is devoted to the so-called needle pulse, which is, in a sense, a three-dimensional generalization of the wave packet studied in the previous section as its intensity distribution resembles a tube radially expanding from the symmetry axis with time. It is an interesting model object because some well-known superluminal propagation-invariant non-diffracting space-time packets can be derived from its simple factorized wave function.

Finally, in Section 3.4, we study the backflow in an *unidirectional modification* of the so-called focus wave mode (FWM)—a pulse propagating luminally (i.e., its group velocity is $c$ in empty space) and without spread, which has been the first and most studied representative of the (3+1)D non-diffracting propagation-invariant space–time wave packets. The FWM possesses a simple closed-form expression but has considerable contributions from backward propagating ('non-causal') plane wave constituents, and is therefore non-interesting from the point of view of backflow study. Fortunately, although not simple, a closed-form expression for a unidirectional version of FWM has been found [47], which has a remarkable backflow as shown further.

## 2. Methods

The essence of our method to study the energy flow is calculation of the spatio-temporal distribution of the Poynting vector **P** and the energy density $u$. For that, in the case of complex scalar fields, we use the following expressions [48]:

$$\mathbf{P} = -\frac{1}{2}\left(\partial_{ct}\Psi^*\right)(\nabla\Psi) - \frac{1}{2}\left(\partial_{ct}\Psi\right)(\nabla\Psi^*) , \tag{1}$$

$$u = \frac{1}{2}\left(\partial_{ct}\Psi\right)\left(\partial_{ct}^*\Psi\right) + \frac{1}{2}(\nabla\Psi)\cdot(\nabla\Psi^*) . \tag{2}$$

For real-valued fields, Equations (1) and (2) reduce to

$$\mathbf{P} = -\frac{\partial}{\partial ct}\psi\,\nabla\psi, \tag{3}$$

$$u = \frac{1}{2}\left(\frac{\partial}{\partial ct}\psi\right)^2 + \frac{1}{2}\nabla\psi\cdot\nabla\psi. \tag{4}$$

For vector-valued fields, the Poynting vector and the energy density are calculated by well-known expressions of electrodynamics [44]. Shortly, a wave function of a scalar field is ascribed to a component (typically to the $z$-component) of the Hertz vector $\mathbf{\Pi}$, from which the Riemann–Silberstein vector

$$\mathbf{F} = \nabla \times \nabla \times \mathbf{\Pi} + (i/c)\,\nabla \times (\partial\mathbf{\Pi}/\partial t) \tag{5}$$

is obtained. In turn, the latter is related to the Poynting vector as $\mathbf{P} = \mathbf{E} \times \mathbf{H} = -i\mathbf{F}^* \times \mathbf{F}$ and the energy density $u = \mathbf{F}^* \cdot \mathbf{F}$. The energy velocity is defined as $\mathbf{V} = \mathbf{P}/u$.

For calculations, we use various packages of scientific numerical and symbolic computations: Mathematica 13.1, Maple 2022.2, Scientific Workplace 5.50, and Mathcad 15 not only to find the best package for every task but also to cross-check the results.

As checked by different scientific computing packages, all studied solutions obey the homogeneous free-space wave equation, propagate without singularities or unusual asymptotic behaviors, and have finite energy density. Some of the solutions are quite complex for analytical proofs, and in these cases we used numerical computation engines. In this way, we were able to consider even quantities dependent on space-integrated total energies, as described in Section 3, if the latter are infinite as it is common for propagation-invariant space–time waves.

It should be pointed out that although calculations based on the closed-form expressions for the wavefields constitute a basic method in the present study, it does not mean that the results obtained hold only for such wavefields (which are mostly impossible or at best difficult to generate experimentally). Correspondingly, the designed numerically presentable and experimentally feasible wavefields may possess the relevant properties analogous to those of their analytical examples (see discussion of this point, e.g., in Ref. [34]).

## 3. Results and Discussion

### 3.1. A (2+1)-Dimensional Unidirectional Monochromatic Wave

In this section, we consider two-dimensional monochromatic waves of frequency $\omega = ck$, and therefore start with a general solution to the 2D Helmholtz equation independent of the $y$-coordinate

$$W(x,z) = \int dk_x e^{-ik_x x} \quad (6)$$
$$\times \int dk_z e^{-ik_z z} \delta(k_x^2 + k_z^2 - \frac{\omega^2}{c^2}) G(k_x, k_z)$$
$$= \int_{-\omega/c}^{\omega/c} dk_z e^{-ik_z z} \frac{\cos Kx}{K} G(k_z).$$

Here, $k_x$ and $k_z$ are corresponding components of the wave vector **k,** and $\delta$ is the Dirac delta function which takes into account the dispersion relation. Using the identity $\delta(a^2 - b^2) = (1/2|b|)[\delta(a-b) + \delta(a+b)]$, resolving and integrating with respect to $k_x$ results in the band-limited Fourier transform of the product $G(k_z)K^{-1}\cos Kx$, where $K = \sqrt{\omega^2/c^2 - k_z^2} \equiv \sqrt{k^2 - k_z^2}$ and $G(k_z)$ is a spectral amplitude function. If $G(k_z) = const = \pi^{-1}$ (for normalization), the Fourier transform is well known and we obtain, finally,

$$W_0(x,z) = J_0\left(\frac{\omega}{c}\sqrt{x^2 + z^2}\right) = J_0(k\rho), \quad (7)$$

where $J_0$ is the the zeroeth order Bessel function of the first kind, and in the last equality, the radial coordinate $\rho = \sqrt{x^2 + z^2}$ of the cylindrical system is introduced. Thus, the field has cylindrical symmetry with respect to the axis $y$. The symmetry results from the omnidirectional angular spectrum.

#### 3.1.1. A New Unidirectional Wave

Our aim is to find an analytic (closed-form) expression of a unidirectional wave. This means that in Equation (6), the condition $G(k_z) = 0$ must be fulfilled for $k_z < 0$ or $G(k_z) = \theta(k_z)\theta(k - k_z)G'(k_z)$, where $\theta(.)$ is the Heaviside unit step function. The only closed-form expression for the integral with a unidirectional spectrum we could find was a combination of cosine and sine transforms with respect to $k_z$ in [49], Section 1.7, Equation (46) and Section 2.7,

Equation (32), viz.,

$$W(x, z) = \mathcal{F}_{\cos}\left\{\theta(k - k_z)\frac{\sqrt{k}\cos\left(x\sqrt{k^2 - k_z^2}\right)}{\sqrt{k_z}\sqrt{k^2 - k_z^2}}\right\} \qquad (8)$$

$$+ i\mathcal{F}_{\sin}\left\{\theta(k - k_z)\frac{\sqrt{k}\cos\left(x\sqrt{k^2 - k_z^2}\right)}{\sqrt{k_z}\sqrt{k^2 - k_z^2}}\right\},$$

which yields the following expression with Bessel functions of fractional orders:

$$W(x, z) = C\sqrt{k\ |z|}\left\{\begin{array}{c} J_{-\frac{1}{4}}\left[\frac{k}{2}\left(\sqrt{x^2 + z^2} - x\right)\right] \\ \times J_{-\frac{1}{4}}\left[\frac{k}{2}\left(\sqrt{x^2 + z^2} + x\right)\right] \\ +i\ \text{signum}\ (z) \\ \times J_{\frac{1}{4}}\left[\frac{k}{2}\left(\sqrt{x^2 + z^2} - x\right)\right] \\ \times J_{\frac{1}{4}}\left[\frac{k}{2}\left(\sqrt{x^2 + z^2} + x\right)\right] \end{array}\right\}. \qquad (9)$$

Here, in order to ensure normalization of the field to unity at the origin, we introduced the constant $C = \Gamma(3/4)^2/2 \approx 0.75$, where $\Gamma$ is the gamma function. The factor $\sqrt{k}$ was introduced into the spectrum in Equation (8) for making $W(x, z)$ dimensionless. It is important to notice that as the arguments of the Bessel functions are non-negative, they have no imaginary parts, and a simple complex conjugation symmetry holds $W(x, z < 0) = W^*(x, z > 0)$ in accordance with Equation (6) if the lower integration limit is set to zero and $G(k_z)$ is a real-valued function. Let us stress that the spectrum was selected not only based on whether a table integral for it exists, but first of all on the unidirectionality condition. For example, Gaussian-type spectra were excluded due to their tails extending to $\pm\infty$.

Since $J_{-\frac{1}{4}}(z)$ diverges at $z = 0$, Equation (9) is not suitable for expressing the field along the axis $x$. Instead, we evaluated the corresponding integral

$$W(x, 0) = \sqrt{k}\int_0^k dk_z \frac{\cos Kx}{\sqrt{k_z}K} = \int_0^{\pi/2} d\phi \frac{\cos(k\,x\sin\phi)}{\sqrt{\cos\phi}}. \qquad (10)$$

The second integral results from the variable transformation $k_z = k\cos\phi$. We see that the angular spectrum, or the density of directions of the plane-wave constituents, varies as $1/\sqrt{\cos\phi}$ and is infinitely strong for the plane waves falling perpendicularly to the axis $z$ while the integral of the spectrum is finite. This is analogous to the case of the field of two superposed plane waves falling under the angle $\phi_0$ onto the axis $z$ when the spectrum is $\delta(\phi - \phi_0)$, i.e., diverging, but the field turns out to be finite. The other diverging factor in Equation (6), $K^{-1} = \left(k^2 - k_z^2\right)^{-1/2} = (k\sin\phi)^{-1/2}$, takes into account the infinite growth of density of plane wave directions per $dk_z$ if $k_z \to k$, but it cancels out in the change in the integration variable to $\phi$.

The last integral in Equation (10) has the following analytic solution:

$$W(x, 0) = C\left(\frac{k\,x}{2}\right)^{\frac{1}{4}}\frac{3J_{\frac{3}{4}}(k\,x) - 2k\,x\,J_{\frac{7}{4}}(k\,x)}{\Gamma\left(\frac{3}{4}\right)k\,x}. \qquad (11)$$

We evaluated and checked the correctness of Equations (9) and (11) with the help of three different software packages for scientific computations, also by direct numerical integration of Equation (8), as well as by subjecting the field to the Helmholtz equation.

The modulus squared of the wave function given by Equations (9) and (11) is plotted in Figure 1. As it is obvious from Equation (9), the real part of $W(x,z)$ is even and the imaginary part is odd with respect to the variable $z$, and both are even with respect to $x$. Therefore, $W(x,0)$ must be also even with respect to $x$ and purely real irrespective of the sign of $x$. The latter features are not directly seen from Equation (11) but an analysis proves their trueness. To the best of our knowledge, the unidirectional 2D monochromatic wave function given in Equation (9) is studied here for the first time.

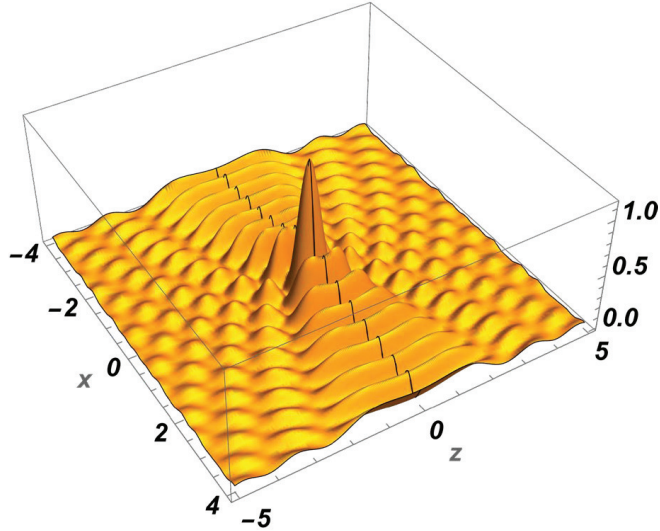

Fig. 1. Plot of the modulus squared $|W(x,z)|^2$ from Equation (9) with $k = 2\pi$. The black line corresponds to $|W(x,0)|^2$.

### 3.1.2. Backflow in the Unidirectional (2+1)-Dimensional Wave

To obtain a full space–time wave function, the time exponential factor $\exp(-ikct)$ must be added to the spatial functions. In the case of Equation (7), this results in a cylindrical standing wave. However, the expression in Equation (9) is a complex-valued function, i.e., it contains a non-zero phase which with the time exponent results in a propagating phase. We found the phase velocity along the axis $z$ by evaluating $\Phi(z) = \arctan\left[\operatorname{Im} W(0,z)/\operatorname{Re} W(0,z)\right]$. The phase $\Phi(z)$ turned out to be in good approximation linear with respect to $z$: the first term in the series expansion at $z = 0$ is $2\pi^{-2}\Gamma(3/4)^4\, kz \approx 0.457\, kz$ while the next term is $-0.0013\,(k z)^3$, i.e., much smaller at least if $kz \leq 2\pi$. Hence, the phase velocity $v_{ph} \approx 2.2c$ and this value does not depend on the value of the wave vector $k$. This can be understood if one notices that in the case of angular spectrum $(\cos\phi)^{-1/2}$ the average angle is about $1\,rad$ and a plane wave falling onto the $z$-axis under such angle has the superluminal phase velocity just about $2c$ along the $z$-axis.

Our unidirectional wave undergoes propagation and it makes sense to calculate the energy flow using Equations (1) and (2). For the field $\Psi(x,z,t) = W(x,z)\exp(-ikct)$, the $z$-component of the Poynting vector is shown in Figure 2.

The dimensionless wave number $k = \pi$ and correspondingly the length unit is half of the wavelength $\lambda = 2\pi/k = 2$. The time instant $ct = 0.5$. However, in distinction from the case of the real versions of the field $\operatorname{Re}\{\Psi(x,z,t)\}$ and $\operatorname{Im}\{\Psi(x,z,t)\}$, neither the Poynting vector nor the energy density of the complex-valued field $\Psi(x,z,t)$ depend on time, as is true of monochromatic wave fields in general. The dimensionless coordinates allow to adopt easily the plots for optical, microwave, or acoustic frequency ranges. For example, if we took the length unit equal to 0.25 µm, the wavelength would be 0.5 µm, which is in the middle of the optical region, and the time instant $ct = 0.5$ would correspond to $t \approx 0.4$ femtosec.

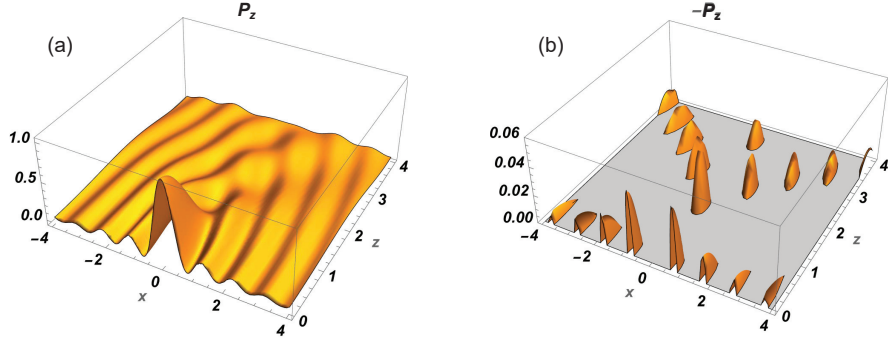

Fig. 2. Spatial dependence of the $z$-component of the Poynting vector of the complex-valued field $\Psi(x, z, t) = W(x, z) \exp(-ikct)$: (**a**) for the $z$-component of the Poynting vector, (**b**) the same for its negative values (backflow) which are depicted with reversed sign and blocked positive values of $P_z$. The wave number $k = \pi$. Vertical axes are normalized to unity at maximum of $P_z$. Notice that, unlike Figure 1, the plots cover only the positive $z$-axis.

The values of parameters as well as the range of the coordinates here and in subsequent figures were chosen so that the backflow effect and its properties are revealed more or less the best.

The plots in Figures 2 and 3 demonstrate remarkable backflow—up to 1 : 20 relative to the maximal forward flow. As Figure 3a shows, in the backflow maximum region with coordinates $(0, 1.6)$, the backflow velocity reaches about 80% of $c$. It seems to be a rule that the backflow is located in regions of low energy density in proximity of the forward flow maxima. We omit the plots here, but the same holds for wavefields $\text{Re}\{\Psi(x, z, t)\}$ and $\text{Im}\{\Psi(x, z, t)\}$, the backflow being even somewhat stronger in the latter. The Poynting vector and energy density of a complex-valued field are vectorial sums of those of its real and imaginary constituent fields. Consequently, if a region of backflow in one of the constituent fields overlaps with a region of the forward flow on the other constituent, mutual cancellation takes place and weaker backflow may disappear.

We also studied the backflow in the vector-valued version of the field $\Psi(x, z, t)$ derived by the Hertz and Riemann–Silberstein vectors as described in Section 2 and in more detail in ref. [44]. A stronger backflow can be noticed comparing Figures 2b and 4b. This is in agreement with earlier observation that polarization may enhance the backflow [7,8,22]. We draw attention to the vortices of the velocity vector field in the right corners of Figure 5a on the slopes of the energy density distribution. Of course, energy velocity arrows need not to be normal to the energy density contours because energy velocity is not the gradient of the energy density. Although specific results are not incorporated in this article, an examination of the Poynting vectors associated individually with pure TE and TM fields obtained from the scalar wave $\Psi(x, z, t)$ also shows the presence of energy backflow. This is altogether different from the cases of the superposition of four plane waves [5–8] and Bessel beams [19, 20] which require a superposition of TE and TM fields for the appearance of energy backflow.

### 3.2. A (2+1)-Dimensional Unidirectional Wave Packet

Closed-form analytical expressions for pulsed unidirectional waves seem relatively difficult to obtain by using an appropriate superposition of the (2+1)D solution in Equation (9) over the wave number $k$. The same can be said of Lekner's (3+1)D monochromatic unidirectional solution [50]. However, this does not mean that (2+1)D or (3+1)D spatiotemporally localized unidirectional closed-form solutions do not exist.

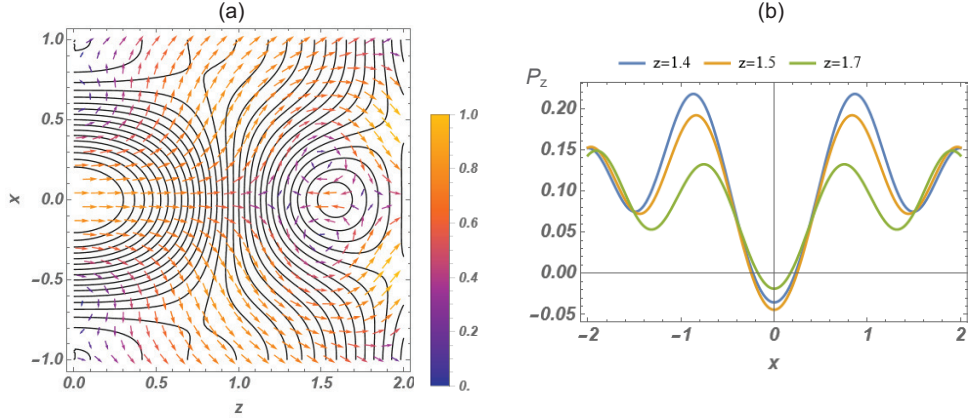

Fig. 3. (**a**) Contour plot of the energy density of the complex-valued field $\Psi(x, z, t) = W(x, z)\exp(-ikct)$ according to Equation (2) superimposed by the vector field plot ($V_x$ and $V_z$ components) of the energy velocity **V**, the color bar showing values of the velocity in the range $0 \ldots c$, where $c = 1$ is the speed of light (or sound in the case of acoustical wave). (**b**) Two-dimensional intersections of the plot Figure 2a of the $z$-component of the Poynting vector.

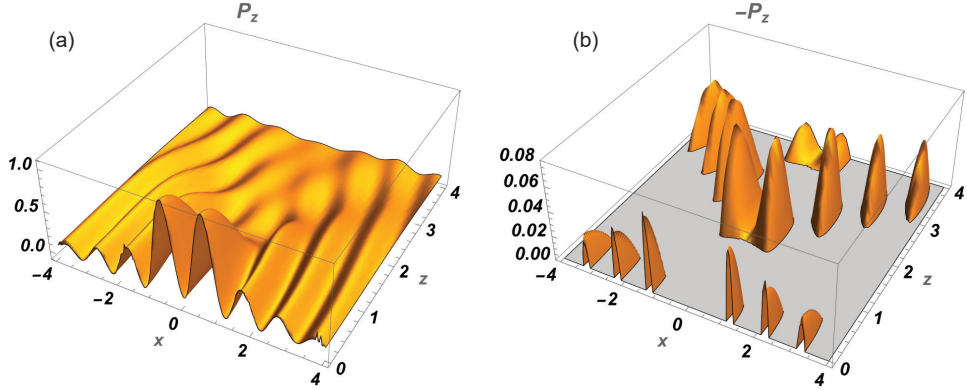

Fig. 4. (**a**) Spatial dependence of the $z$-component of the Poynting vector of the electromagnetic field obtained from $\Psi(x, z, t)$; (**b**) the same for its negative values (backflow). See also caption of Figure 2.

Indeed, the monochromatic (2+1)D field $\Psi(x, z, t) = W(x, z)\exp(-ikct)$ can be integrated over $k$ using the spectrum $\exp(-ak)\sqrt{a/k}$, where $a > 0$ is the pulsewith parameter, with the help of the table integral 6.612-3 in [51]. Let us note that in order to obtain the right results by the table expression with the Legendre function, the branch cut line for the latter must not be placed outside the segment $(-1, 1)$ of the real axis, which is, however, the common placing, e.g., in WolframAlpha on-line calculator and the default one in the Mathematica package. The result turns out to be a closed-form expression with Legendre functions of the second kind $Q_\nu(.)$, and

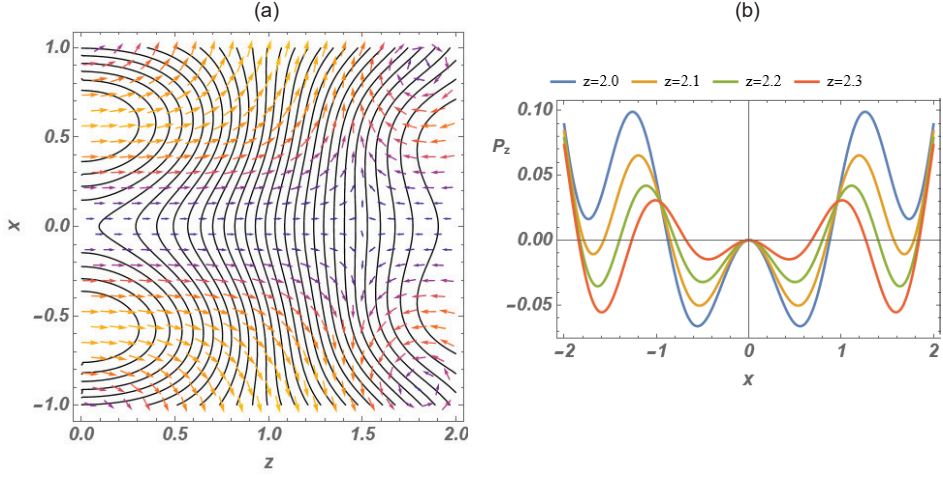

Fig. 5. (**a**) Contour plot of the energy density $\vec{F} \cdot \vec{F}^*$ calculated from $\Psi(x, z, t)$ and superimposed on the vector field plot of the energy velocity. (**b**) Two-dimensional intersections of the plot in Figure 4a of the $z$-component of the Poynting vector.

for $z > 0$ it reads

$$\Phi(x, z, t) = \sqrt{\frac{\pi a}{2z}}\, Q_{-\frac{3}{4}}\left(2z^{-2}\left[x^2 + \frac{1}{2}z^2 + (a + ict)^2\right]\right) \\ + i\sqrt{\frac{\pi a}{2z}}\, Q_{-\frac{1}{4}}\left(2z^{-2}\left[x^2 + \frac{1}{2}z^2 + (a + ict)^2\right]\right). \tag{12}$$

The special functions $Q_\nu(.)$ with complicated arguments make the solution rather cumbersome, but still, it is analytical and allows taking derivatives needed for calculation of the Poynting vector, as well as electromagnetic fields. To the best of our knowledge, the expression in Equation (12) of a (2+1)D unidirectional wave packet is published here for the first time. In [52], another example of a (2+1)D unidirectional wave packet was derived.

At instant $t = 0$, the pulse is narrow in the $x$-direction; see Figure 6a. With time, it acquires a crater-like shape, the radius of the nearly circular edge of which increases with the speed of light; see Figure 6b. The energy density behaves the same way. Figure 6 as well as our other plots made for a series of time instants show that the wave packet is not propagation-invariant and its characteristic feature is two wavefronts which are parallel to the axis $z$ and are running into each other (at $t < 0$) and thereafter (at $t > 0$) running apart (with velocity $c$).

The $z$-component of Poynting vector evaluated from Equation (12) according to Equation (3) is depicted in Figure 7. The forward energy flow is the strongest along the edge of the 'crater'. The backflow is of the same order of magnitude as in the case of the monochromatic field and its maxima are located near the 'notches' in the 'crater' in Figure 6b. The arc of the backflow maxima in Figure 7b has a radius of about nine units and that of the forward flow maxima in Figure 7a has the radius of about eight units. This is in accordance with the expansion of the 'crater' of Figure 6b for the time moment of $ct = 8$ and indicates that the strongest backflow takes place on the front side of the forward flow arc.

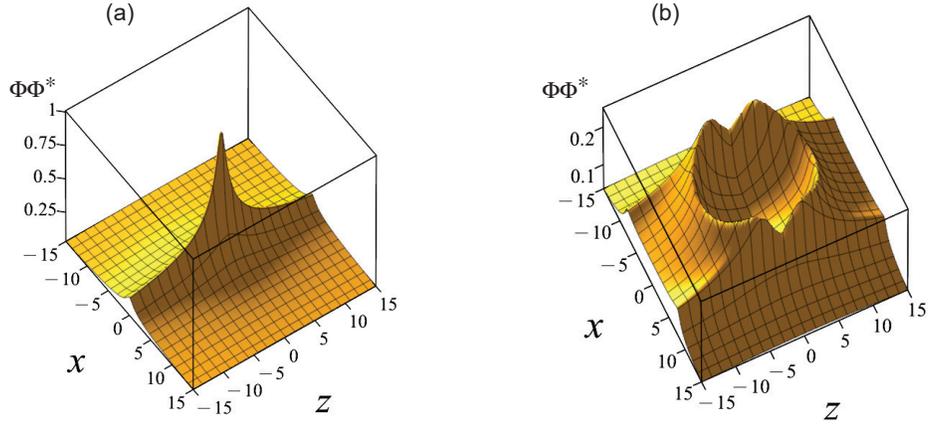

Fig. 6. Square modulus of the wave packet of Equation (12) $a = 1/2$ at (**a**) $ct = 0$ and (**b**) $ct = 8$.

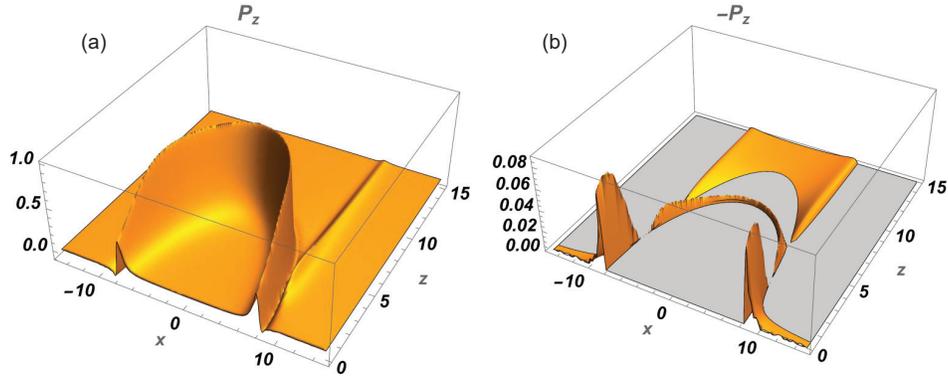

Fig. 7. (**a**) Spatial dependence of the $z$-component of the Poynting vector of the real part of field $\Phi(x, z, t)$. (**b**) The same for its negative values (see also caption of Figure 2b). Notice that unlike Figure 6, the plots cover only the positive $z$-axis. $ct = 8$, $a = 1/2$.

### 3.3. A (3+1)-Dimensional Scalar and Vector-Valued "Needle" Pulse

In a sense, the (3+1)D needle pulse is the simplest and most interesting pulsed wave packet for studying the backflow because it is singularly unidirectional—the $z$-component of wave vectors of its plane wave (or Bessel-beam-) constituents has only one fixed value $k_z = k_{z0} > 0$.

The axisymmetric scalar needle pulse has been derived under different names by several authors; see Equation (4.3) in [28], Equation (13) in [31], and Equation (18) in [52]. Parenthetically, needle pulses studied here must be distinguished from 'pulsed needle Bessel beams' considered in [53]. In cylindrical coordinates $(\rho, z)$, the needle pulse is given by the following expression:

$$\Pi(\rho, z, t) = C \frac{\exp\left(-k_{z0}\sqrt{\rho^2 + (\Delta + ict)^2}\right)}{\sqrt{\rho^2 + (\Delta + ict)^2}} e^{ik_{z0}z}, \tag{13}$$

where $C = \Delta \exp(-k_{z0}\Delta)$ is a normalization constant in order to have $\Pi(0,0,0) = 1$ at the origin and $\Delta > 0$ is a pulsewidth parameter. It describes a simple cylindrical pulse modulated harmonically in the axial direction, which for times $t < 0$ converges radially to the axis $z$ and thereafter for $t > 0$ expands from it, the intensity distribution resembling an infinitely long tube coaxial with the $z$-axis and with time-dependent diameter (see plots in Figure 6 of ref. [31]).

Let it be emphasized that although the needle pulse is arbitrarily strongly non-paraxial, it can be practically generated from a femtosecond laser pulse by cylindrical or circular gratings and conical optics [54, 55]. Similarly, a (2+1)D version of the needle pulse can be realized quite trivially by two planar gratings.

The energy backflow in the needle pulse is depicted in Figure 8. The plots are periodic along the axis $z$ with period equal to 4. This is understandable since the period of the $z$-dependence in Equation (13) with $k_{z0} = \pi/4$ is $2\pi/k_{z0} = 8$, which in the Poynting vector as a quadratic quantity should be two times smaller. At $t = 0$, the backflow is absent and only narrow on-axis maxima of the forward flow are present at locations $z = -4, 0, 4$, etc. For times in the interval $0 < ct < 1$, the backflow maxima located along the axis $z$ prevail and their intensity reaches the fraction 1:7 relative to that of the forward flow maxima. At $ct = 1$, the energy backward flows take place on rings of radii $\rho \approx 0.7$ and $\rho \approx 1.2$ which are coaxial with the forward flow rings of radius $\rho = ct = 1$. The peak values of the intensities of the backflow rings are equal and constitute a factor 1:20 of that of the forward flow (see Figure 8b). As time increases ($ct > 1$), the intensity of the backflow in the inner rings gradually vanishes while that of the outer rings maintains the ratio of $\approx$1:20 relative to the forward flow peak values. It should be mentioned that while the group velocity of the needle pulse is infinitely large [31], the energy velocity is limited by $c$ as it has to be.

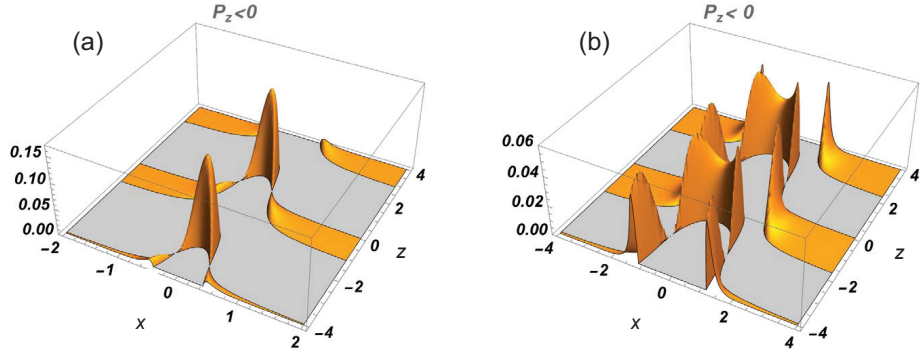

Fig. 8. Spatial dependence of the negative values (backflow) of the $z$-component of the Poynting vector of the real part of $\Pi(\rho, z, t)$ in Equation (13): (**a**) at time instant $ct = 0.25$ and (**b**) at time instant $ct = 1$. Positive values of the $z$-component of the Poynting vector are blocked (see caption of Figure 2b). The axis $x = \pm\rho$ shows any transverse coordinate, and parameter values are $k_{z0} = \pi/4$, $\Delta = 0.2$.

The imaginary part of $\Pi(\rho, z, t)$ leads to the same results, except for a phase shift along the axis $z$.

Of greater physical interest is, of course, the ratio of integral values of the backward and forward flows of energy, i.e., the ratio of total flows. But in evaluation of the integral flows, we encounter an issue: since the field in Equation (13) has infinite energy, the integral of the $z$-component of its Poynting vector along the axis $z$ diverges. Fortunately, thanks to the periodic behavior of the $z$-component, integration over the axis $z$ within limits $z = -\infty ... + \infty$ can be replaced by finite limits when the *ratio* of the backward and forward integral flows needs to be found (provided that a number of full periods is involved in the integration). Parenthetically, a real needle pulse formed by a finite-length cylindrical grating also has a correspondingly finite length along the axis $z$. The ratio of backward and forward energy flows found in such a way turns out to well represent that for the physically realizable needle pulse, as shown below.

We studied along the same lines a *finite-energy* version of a needle pulse. The simplest expression for such pulse can be easily obtained by replacing in Equation (13) $k_{z0}$ with the variable wavenumber $k_z$ and integrating over the wave number with an exponential unidirectional spectrum $\exp(-k_z z_s)\theta(k_z - k_{z0})$, where $z_s$ is a positive constant and $\theta(.)$ is the Heaviside function. The result is

$$F(\rho, z, t) = C_f \frac{\exp\left(-k_{z0}\sqrt{\rho^2 + (\Delta + ict)^2}\right)}{\sqrt{\rho^2 + (\Delta + ict)^2}\left(\sqrt{\rho^2 + (\Delta + ict)^2} - i(z + iz_s)\right)} e^{ik_{z0}(z+iz_s)}, \qquad (14)$$

where $C_f = \Delta(\Delta + z_s)\exp[k_{z0}(\Delta + z_s)]$ is the normalization constant. The second factor in the denominator ensures vanishing of $F(\rho, z, t)$ as well as of the Poynting vector if $z \to \pm\infty$. Let us note that in a more general expression, this factor is raised to power $q > 1$. Let us stress that $F(\rho, z, t)$ is unidirectional despite its generally wide exponentially decaying $k_z$-spectrum. Another finite-energy needle pulse has been studied in [52], where the tail of the Gaussian $k_z$-spectrum in principle involves a small portion of backward-propagating plane wave constituents, thus making the pulse not purely unidirectional.

The distinguishing feature of the field in Equation (14) from that in Equation (13) is its decaying intensity with growth of coordinate $z$ instead of the infinitely periodic behavior. The same holds for the $z$-component of the Poynting vector, so that for $z_s = 2$ the maxima near $z = \pm 4$ (see Figure 8) become weak. For $z_s = 1$ they almost disappear, and for $z_s = 0.1$ practically only the central maximum survives. The same holds for the energy density; see Figure 9a in which the region around the survived maximum is presented with high spatial resolution.

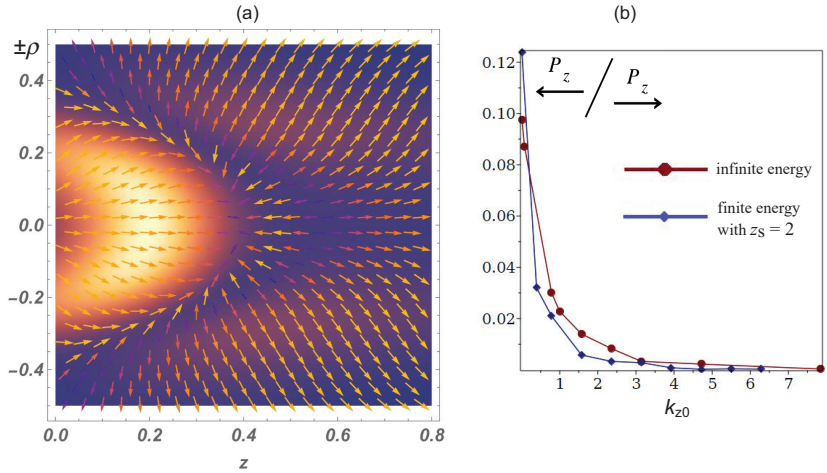

Fig. 9. (**a**) Vector field plot of the energy flow velocity, superimposed on the energy density pseudocolor plot; the parameter values are the same as in Figure 8a with the additional parameter value $z_s = 0.1$. Velocity arrows have color in accordance with the color bar scale shown in Figure 3a. (**b**) Dependence of the ratio $P_z^{\leftarrow}/P_z^{\rightarrow}$ of total backward and forward energy flows on the wave vector $k_{z0}$ in the case of finite-energy needle pulse given by Equation (14) compared to that of the infinite-energy pulse in Equation (13); parameter values $ct = 1$, $\Delta = 0.2$.

The energy flow velocity plot in the region of that maximum clearly shows the presence of backward flow behind the forward flow maximum. It is remarkable that the velocity of the backward flow reaches the value $c$ on the axis $z$ like the velocity of the forward flow.

As to the total (integrated over the whole space) flow, it is sufficient to consider only the $z$-component of the Poynting vector as, due to the cylindrical symmetry, the radial contributions to the total flows vanish. From Figure 9b, one can draw three conclusions. We see that the total backflow decreases with the increase in $k_{z0}$. This is understandable, as the larger $k_{z0}$ is, the smaller is the mean angle of wave vectors with respect to axis $z$ in the wave packet. Indeed, it is known already from the study of a quartet of plane waves [7,8] that the more the wave packet is non-paraxial, the stronger the backflow effect. Second, the curves of the dependence are rather similar, which justifies the aforementioned method of calculation of the ratio of the total flows in the case of the infinite-energy pulse. Third, the ratio of the flows still decreases more rapidly in the case of the finite-energy pulse. This can be explained by the circumstance that the spectrum of the wave given by Equation (14) involves two exponents, $\exp(-z_s k_z)$ and $\exp(-k\Delta)$, while the wave Equation (13) involves only the latter. For very small values $k_{z0} \sim 10^{-3}$, the ratio for the finite-energy pulse with parameters of Figure 9a reaches a value of about 1:5.

A rather strong backflow in the needle pulse as compared to the effect in other wave fields studied here and in papers cited in the Introduction is at first glance difficult to understand. Indeed, the spectrum $\exp(-k_z z_s)\theta(k_z - k_{z0})$ means that in the wave packet the plane wave (or Bessel-beam) constituents with $k_z \approx k_{z0}$ are the strongest. But these constituents have radial components of their wave vectors $k_\rho \approx 0$, i.e., the most paraxial constituents prevail in the wave packet. However, we evaluated the *mean* wave vector in the packet with parameter values $\Delta = 0.2$, $k_{z0} = \pi/4$ and found it to form an angle $\approx 80°$, i.e., the packet is strongly non-paraxial.

We also studied vector-valued versions of the needle pulse obtained with the help of the Hertz and Riemann–Silberstein vectors; see Equation (5). A hope was that vectorization, especially involvement of combined TE and TM polarizations, would enhance the backflow effect. Indeed, it is known that, e.g., only vector-valued versions of the zeroth-order Bessel beams and X waves exhibit the effect and the TE+TM polarization is preferred [20–22]. However, none of the studied versions of the vector-valued needle pulse demonstrated remarkably stronger backflow than the scalar versions.

It should be noted, finally, that the infinite-energy needle pulse is not a propagation- invariant space–time wavepacket because its modulus is independent of the axial coordinate $z$. However, if observed from another inertial frame moving with a relativistic velocity along the axis [28, 31] (or generated in a laboratory with special gratings [54]), the needle pulse transforms into a propagation-invariant space–time wave packet called the focused X wave (FXW), whose modulus depends solely on variable $z - vt$ where $v$ is superluminal velocity. We omit results on the FXW because it exhibits very weak backflow, practically disappearing if $v < 10c$. In turn, if $k_{z0} = 0$, the FXW transforms into the superluminal axisymmetric zeroth-order X wave [25–29, 31, 32, 37–39] which completely lacks the backflow effect.

### 3.4. A (3+1)D Unidirectional Version of the Luminal Propagation-Invariant space–time Wavepacket—The Focus Wave Mode (FWM)

The rather complicated closed-form expression for a *unidirectional* version of FWM, which contains Lommel functions of two arguments, has been derived in [47]. The field of the pulse can be written in cylindrical coordinates as the following difference:

$$U_{k_1,\infty}(\rho, z, t) = U_{k_0,\infty}(\rho, z, t) - U_{k_0,k_1}(\rho, z, t) . \tag{15}$$

Here, $U_{k_0,\infty}(\rho, z, t)$ is the well-known expression of the FWM which is obtained by superposing all Bessel-beam constituents with wavenumbers $k$ from the minimal value $k_0$ up to infinity. This means that in the superposition, the corresponding integration for the $z$-component of the wavenumber runs over the range $k_z \in (-k_0, \infty)$, i.e, the FWM contains considerable contribution from the backward-propagating constituents. Equation (15) states that if we substract from the FWM, a field obtained as a superposition of the constituents with $k$ from $k_0$ up to a value $k_1$ which corresponds to nonnegative $k_z$, we obtain the expression of the unidirectional field $U_{k_1,\infty}(\rho,z,t)$ we are looking for. The subtracted field $U_{k_0,k_1}(\rho,z,t)$ is the one which is expressed in terms of Lommel functions. Choosing $k_1 = 2k_0$, the integration for the $z$-component runs over the range $k_z \in (0,\infty)$, i.e., the difference field $U_{k_1,\infty}(\rho,z,t)$ contains all forward-progating Bessel-beam constituents [47].

Animated 3D plots of space–time dependence of the field $U_{k_1,\infty}(\rho,z,t)$ can be seen in Supplementary Material of ref. [47]. The dependence of the $z$-component of the Poynting vector, depicted in Figure 10a, is similar to that of the modulus squared of $U_{k_1,\infty}(\rho,z,t)$.

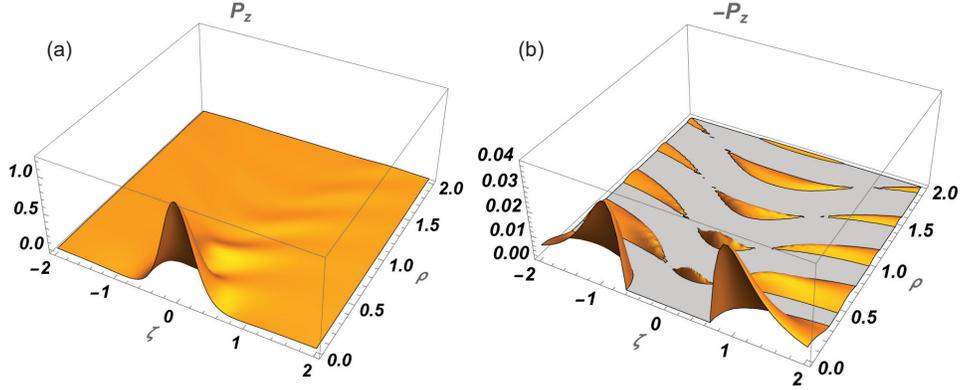

Fig. 10. (**a**) Spatial dependence of the $z$-component of the Poynting vector of $\mathrm{Im}\{U_{k_1,\infty}(\rho,z,t)\}$. (**b**) The same for its negative values (backflow). $\zeta \equiv z - ct$, $k_0 = \pi$, $k_1 = 2\pi$, $z_0 = 2/k_0$, $c = 1$, $t = 0$. Notice that the orientation of the 3D plots differs from that of preceeding figures.

The frequency or wave number spectrum of the FWM is determined by a decaying exponential $\exp(-z_0 k)$. While the support of the spectrum of the needle pulse in the plane $(k_z, \rho)$ is a vertical line $k_z = k_{z0}$ and therefore the strength of spectrum does not depend on $k_z$, the support of the FWM in the plane $(k_z, \rho)$ is a parabola [47], along the arm of which the strength of the spectrum decays as $\exp(-z_0 k_z)$ with $0 < k_z \to \infty$. Hence, contrary to the needle pulse, the plane-wave constituents with wave vectors perpendicular to the axis $z$ contribute most strongly into the FWM, and the smaller the decay parameter $z_0$, the smaller the angle between the average wave vector of the constituents and axis $z$. In other words, contrary to the needle pulse, increasing $z_0$ results in enhancement of the backflow in the FMW. Indeed, that is the case: while in the case $z_0 = 2/k_0$ the ratio of peak value of the backflow in Figure 10b to that of the forward flow in Figure 10a is 1:35, increasing $z_0$ threefold leads to the ratio of 1:29. The angles between the constituent wave vectors and the $z$-axis averaged over the spectrum are 78° and 85°, i.e., increasing $z_0$ enhances the non-paraxiality.

Backflow peaks in the field $\mathrm{Re}\{U_{k_1,\infty}(\rho,z,t)\}$ are by an order of magnitude weaker at $t = 0$ but become pretty much similar to those in Figure 10 for a shifted time instant $ct = \pm 0.25$ which corresponds to $\pm 1/4$ of the period $T = 2\pi/ck_1$. This is understandable, since such a quarter-cycle temporal shift not only moves the pulse along the $z$-axis with velocity $c$ but also transforms the field $\mathrm{Re}\{U_{k_1,\infty}(\rho,z,t)\}$ approximately into a shifted $\mathrm{Im}\{U_{k_1,\infty}(\rho,z,t)\}$ and vice versa.

## 4. Conclusions

Wave functions formed from plane wave constituents—or Bessel beam ones in the case of cylindrical 3D waves—propagating only in the positive, say $z$-direction, are called unidirectional. If they contain finite energy, they can be launched causally from an aperture. Even if they have theoretically infinite energy, in reality, a finite-size aperture results in their finite-energy approximations which differ from the theoretical wave functions mainly by off-axis edge effects. It is for unidirectional wave functions that the very question of energy backflow is meaningful.

The present work together with the results of previous relevant studies allows us to draw the following general conclusions.

1. The energy backflow is a weak effect and typically constitutes a fraction less than 10% from the forward flow.

2. Nevertheless, the velocity of the energy backflow may reach the physical maximum $c$.

3. Non-paraxiality enhances the effect. However, it is not a sufficient condition—e.g., the X-wave can be extremely non-paraxial but still absolutely lacking the backflow.

4. The vector nature of a wavefield is by far not always crucial for the appearance of the backflow but generally enhances the effect in comparison with a corresponding scalar-valued wavefield.

5. Stronger backflow tends to be located in regions of low energy density and/or in the vicinity of maxima of the forward flow. Reasons for that need further study.

In summary, the energy backflow is an intriguing effect in the physics of homogeneous waves that propagate in free space without singularities, and it is essential for various physical phenomena in which the direction of the Poynting vector is important.

## References

1. Bialynicki-Birula, I.; Bialynicka-Birula, Z.; Augustynowicz, S. Backflow in relativistic wave equations. *J. Phys. A Math. Theor.* **2022**, *55*, 255702.
2. Bracken, A.J. Probability flow for a free particle: New quantum effects. *Phys. Scr.* **2021**, *96*, 045201.
3. Barbier, M.; Fewster, C.J.; Goussev, A.; Morozov, G.V.; Srivastava, S.C.L. Comment on 'Backflow in relativistic wave equations'. *J. Phys. A Math. Theor.* **2023**, *56*, 138003.
4. Bracken, A.; Melloy, G. *J. Phys. A Math. Theor.* **2023**, *56*, 138002.
5. Bialynicki-Birula, I.; Bialynicka-Birula, Z.; Augustynowicz, S. Reply to comments on 'Backflow in relativistic wave equations'. *J. Phys. A Math. Theor.* **2023**, *56*, 138001.
6. Katsenelenbaum, B. What is the direction of the Poynting vector? (A methodic note). *J. Commun. Technol. Electron.* **1997**, *42*, 119–120.
7. You, X.L.; Li, C.F. From Poynting vector to new degree of freedom of polarization. *arXiv* **2020**, arXiv:2009.04119.
8. Saari, P.; Besieris, I. Backward energy flow in simple four-wave electromagnetic fields. *Eur. J. Phys.* **2021**, *42*, 055301.
9. Ustinov, A.V.; Porfirev, A.P.; Khonina, S.N. Interference Generation of a Reverse Energy Flow with Varying Orbital and Spin Angular Momentum Density. *Photonics* **2024**, *11*, 962.
10. Richards, B.; Wolf, E. Electromagnetic diffraction in optical systems, II. Structure of the image field in an aplanatic system. *Proc. R. Soc. Lond. A* **1959**, *253*, 358–379.
11. Kotlyar, V.; Stafeev, S.; Nalimov, A.; Kovalev, A.; Porfirev, A. Mechanism of formation of an inverse energy flow in a sharp focus. *Phys. Rev. A* **2020**, *101*, 033811.
12. Li, H.; Wang, C.; Tang, M.; Li, X. Controlled negative energy flow in the focus of a radial polarized optical beam. *Opt. Express* **2020**, *28*, 18607–18615.
13. Han, L.; Qi, J.; Gao, C.; Li, F. Controllable reverse energy flow in the focus of tightly focused hybrid vector beams. *Opt. Express* **2024**, *32*, 36865–36874.
14. Geints, Y.E.; Minin, I.V.; Minin, O.V. Simulation of enhanced optical trapping in a perforated dielectric microsphere amplified by resonant energy backflow. *Opt. Commun.* **2022**, *524*, 128779.
15. Yuan, G.; Rogers, E.T.; Zheludev, N.I. "Plasmonics" in free space: Observation of giant wavevectors, vortices, and energy backflow in superoscillatory optical fields. *Light. Sci. Appl.* **2019**, *8*, 2.
16. Lukyanchuk, B.; Wang, Z.; Tribelsky, M.; Ternovsky, V.; Hong, M.; Chong, T. Peculiarities of light scattering by nanoparticles and nanowires near plasmon resonance frequencies. *J. Physics: Conf. Ser.* **2007**, *59*, 234.

17. Tribelsky, M.I.; Luk'yanchuk, B.S. Anomalous light scattering by small particles. *Phys. Rev. Lett.* **2006**, *97*, 263902.
18. Eliezer, Y.; Zacharias, T.; Bahabad, A. Observation of optical backflow. *Optica* **2020**, *7*, 72–76.
19. Ghosh, B.; Daniel, A.; Gorzkowski, B.; Bekshaev, A.Y.; Lapkiewicz, R.; Bliokh, K.Y. Canonical and Poynting currents in propagation and diffraction of structured light: Tutorial. *JOSA B* **2024**, *41*, 1276–1289.
20. Turunen, J.; Friberg, A.T. Self-imaging and propagation-invariance in electromagnetic fields. *Pure Appl. Opt.* **1993**, *2*, 51.
21. Novitsky, A.V.; Novitsky, D.V. Negative propagation of vector Bessel beams. *JOSA A* **2007**, *24*, 2844–2849.
22. Salem, M.A.; Bağcı, H. Energy flow characteristics of vector X-waves. *Opt. Express* **2011**, *19*, 8526–8532.
23. Brittingham, J.N. Focus waves modes in homogeneous Maxwell's equations: Transverse electric mode. *J. Appl. Phys.* **1983**, *54*, 1179–1189.
24. Besieris, I.M.; Shaarawi, A.M.; Ziolkowski, R.W. A bidirectional traveling plane wave representation of exact solutions of the scalar wave equation. *J. Math. Phys.* **1989**, *30*, 1254–1269.
25. Lu, J.Y.; Greenleaf, J.F. Nondiffracting X waves-exact solutions to free-space scalar wave equation and their finite aperture realizations. *IEEE Trans. Ultrason. Ferroelectr. Freq. Control* **1992**, *39*, 19–31.
26. Ziolkowski, R.W.; Besieris, I.M.; Shaarawi, A.M. Aperture realizations of exact solutions to homogeneous-wave equations. *JOSA A* **1993**, *10*, 75–87.
27. Saari, P.; Reivelt, K. Evidence of X-shaped propagation-invariant localized light waves. *Phys. Rev. Lett.* **1997**, *79*, 4135.
28. Besieris, J.; Abdel-Rahman, M.; Shaarawi, A.; Chatzipetros, A. Two fundamental representations of localized pulse solutions to the scalar wave equation. *J. Electromagn. Waves Appl.* **1998**, *12*, 981–984.
29. Salo, J.; Fagerholm, J.; Friberg, A.T.; Salomaa, M. Unified description of nondiffracting X and Y waves. *Phys. Rev. E* **2000**, *62*, 4261.
30. Grunwald, R.; Kebbel, V.; Griebner, U.; Neumann, U.; Kummrow, A.; Rini, M.; Nibbering, E.; Piché, M.; Rousseau, G.; Fortin, M. Generation and characterization of spatially and temporally localized few-cycle optical wave packets. *Phys. Rev. A* **2003**, *67*, 063820.
31. Saari, P.; Reivelt, K. Generation and classification of localized waves by Lorentz transformations in Fourier space. *Phys. Rev. E* **2004**, *69*, 036612.
32. Kiselev, A. Localized light waves: Paraxial and exact solutions of the wave equation (a review). *Opt. Spectrosc.* **2007**, *102*, 603–622.
33. Yessenov, M.; Bhaduri, B.; Kondakci, H.E.; Abouraddy, A.F. Classification of propagation-invariant space–time wave packets in free space: Theory and experiments. *Phys. Rev. A* **2019**, *99*, 023856.
34. Reivelt, K.; Saari, P. Experimental demonstration of realizability of optical focus wave modes. *Phys. Rev. E* **2002**, *66*, 056611.
35. Bowlan, P.; Valtna-Lukner, H.; Lõhmus, M.; Piksarv, P.; Saari, P.; Trebino, R. Measuring the spatiotemporal field of ultrashort Bessel-X pulses. *Opt. Lett.* **2009**, *34*, 2276–2278.
36. Kondakci, H.E.; Abouraddy, A.F. Diffraction-free space–time light sheets. *Nat. Photonics* **2017**, *11*, 733–740.
37. Recami, E.; Zamboni-Rached, M.; Hernández-Figueroa, H.E. (Eds.) *Localized Waves: A Historical and Scientific Introduction*; John Wiley & Sons: Hoboken, NJ, USA, 2008.
38. Hernández-Figueroa, H.E.; Zamboni-Rached, M.; Recami, E. (Eds.) *Non-Diffracting Waves*; John Wiley & Sons: Hoboken, NJ, USA, 2013.
39. Yessenov, M.; Hall, L.A.; Schepler, K.L.; Abouraddy, A.F. space–time wave packets. *Adv. Opt. Photonics* **2022**, *14*, 455–570.
40. Saari, P. Reexamination of group velocities of structured light pulses. *Phys. Rev. A* **2018**, *97*, 063824.
41. Saari, P.; Rebane, O.; Besieris, I. Energy-flow velocities of nondiffracting localized waves. *Phys. Rev. A* **2019**, *100*, 013849.
42. Zamboni-Rached, M. Unidirectional decomposition method for obtaining exact localized wave solutions totally free of backward components. *Phys. Rev. A* **2009**, *79*, 013816.
43. So, I.A.; Plachenov, A.B.; Kiselev, A.P. Simple unidirectional finite-energy pulses. *Phys. Rev. A* **2020**, *102*, 063529.
44. Besieris, I.; Saari, P. Energy backflow in unidirectional spatiotemporally localized wave packets. *Phys. Rev. A* **2023**, *107*, 033502.
45. Lekner, J. Family of oscillatory electromagnetic pulses. *Phys. Rev. A* **2023**, *108*, 063502.
46. Lekner, J. Erratum: Family of oscillatory electromagnetic pulses [Phys. Rev. A 108, 063502 (2023)]. *Phys. Rev. A* **2024**, *109*, 059901.
47. Sheppard, C.J.; Saari, P. Lommel pulses: An analytic form for localized waves of the focus wave mode type with bandlimited spectrum. *Opt. Express* **2008**, *16*, 150–160.
48. Mandel, L.; Wolf, E. *Optical Coherence and Quantum Optics*; Cambridge University Press: Cambridge, UK, 1995; p. 288.
49. Erdelyi, A. *Tables of Integral Transforms*; McGraw-Hill: New York, NY, USA, 1954; Volume 1.
50. Lekner, J. Tight focusing of light beams: A set of exact solutions. *Proc. R. Soc. A* **2016**, *472*, 20160538.
51. Gradshteyn, I.; Ryzhik, I. *Table of Integrals, Series, and Products*, 6th ed.; Academic Press: New York, NY, USA, 2000.
52. Parker, K.J.; Alonso, M.A. Longitudinal iso-phase condition and needle pulses. *Opt. Express* **2016**, *24*, 28669–28677.
53. Grunwald, R.; Bock, M. Needle beams: A review. *Adv. Phys. X* **2020**, *5*, 1736950.

54. Valtna, H.; Reivelt, K.; Saari, P. Methods for generating wideband localized waves of superluminal group velocity. *Opt. Commun.* **2007**, *278*, 1–7.
55. Saari, P.; Reivelt, K.; Valtna, H. Ultralocalized superluminal light pulses. *Laser Phys.* **2007**, *17*, 297–301.